\documentstyle{mn}
\input psfig.sty

\title[binary evolution]
{Low and intermediate-mass close binary evolution and the initial -- final
mass relation}
\author[Z. Han, C.A. Tout, P.P. Eggleton]
{Zhanwen Han$^{1,2,3}$, Christopher A. Tout$^3$, Peter P.
Eggleton$^{3,4}$\\
$^1$Yunnan Observatory, Academia Sinica, Kunming, 650011, P.R.China
(zhanwen@public.km.yn.cn)\\
$^2$National Astronomical Observatories, Chinese Academy of Sciences\\
$^3$Institute of Astronomy, Madingley Road, Cambridge, CB3 0HA, UK\\
$^4$IGPP, L-413, Lawrence Livermore National Laboratory,
     7000 East Avenue,
     Livermore CA, 94550, USA
}

\begin{document}
\maketitle
\begin{abstract}
Using Eggleton's stellar evolution
code, we carry out 150 runs of Pop I binary evolution
calculations, with the initial primary mass between 1 and $8M_\odot$, the
initial mass ratio $q=M_1/M_2$ between 1.1 and 4, and the onset of Roche
lobe
overflow (RLOF) at an early, middle, or late Hertzsprung-gap stage.
We assume that RLOF is conservative in the calculations, and find that
the remnant mass of the primary may change by more than 40 per cent
over the range of initial mass ratio or
orbital period, for a given primary mass.
This is contrary to the often-held belief that the remnant mass depends only
on the
progenitor mass if mass transfer begins in the Hertzsprung gap.
We fit a formula, with an error less than 3.6 per cent,
for the remnant (white dwarf) mass as a function
of the initial mass $M_{\rm 1i}$ of the primary, the initial mass ratio
$q_{\rm i}$, and the radius of the primary at the onset of RLOF.
We also find that a carbon-oxygen white dwarf with mass as low as
$0.33M_\odot$ may be formed if the primary's initial mass is around
$2.5M_\odot$.
\end{abstract}


\section{Introduction}
Binary evolution plays an important role in the formation of many
interesting
stellar objects and binary evolution theory has been successful in
solving many puzzles in the past.
 In particular, double degenerates, as possible supernovae Type Ia progenitors
\cite{ibe84,web84},
are receiving more and more attention.
The confrontation of the
double degenerate theory of
Iben et al.\ \shortcite{ibe97} and Han \shortcite{han98}
with the observations of Marsh's group \cite{mar00}
shows a conflict between the theoretical and
observational mass ratio distribution of double degenerates.
An Algol-like phase is an important or even the main channel for the
formation of double degenerates. The confrontation indicates that
we need a systematic investigation of binary evolution with
the onset of Roche lobe overflow (RLOF) in the Hertzsprung gap (HG).

Case-B binary evolution,
in which mass transfer begins after central hydrogen exhaustion,
has been studied  in the past.
Refsdal \& Weigert \shortcite{ref69} and
Giannone \& Giannuzzi \shortcite{gia70} calculated several runs of
case B evolution for low-mass binaries.
Van der Linden \shortcite{van87} calculated conservative evolution for
initial
primary masses  between 3 and $12M_\odot$.
De Loore \& De Greve \shortcite{del92},
De Greve \& de Loore \shortcite{deg92},
De Greve \shortcite{deg93}
and de Loore \& Vanbeveren \shortcite{del94a,del94b,del95}
carried out non-conservative evolution (conservative evolution in
some cases) for
initial primary masses between 3 and $40M_\odot$.
Sarna et al.\ \shortcite{sar96} investigated evolutionary scenarios
for double degenerates. The evolutionary sequences calculated previously
mostly cover different initial primary masses, initial orbital periods that
make RLOF begin in the early HG, and only one or two initial mass
ratios.
Furthermore there is a lack of low-mass binary sequences.

A binary population synthesis of double degenerates \cite{han98}
requires a detailed knowledge of binary evolution. In this paper,
we investigate binary evolution with the onset of RLOF in the
Hertzsprung gap in a systematic way. We calculate
150 runs of binary evolution with a rather complete parameter space,
i.e. varying the primary's initial mass between 1 and $8M_\odot$, the
initial mass ratio $q=M_1/M_2$ between 1.1 and 4, and the onset of Roche lobe
overflow (RLOF) at early, middle, or late Hertzsprung gap.

\section{Computations}

\begin{figure}
\centerline{\psfig{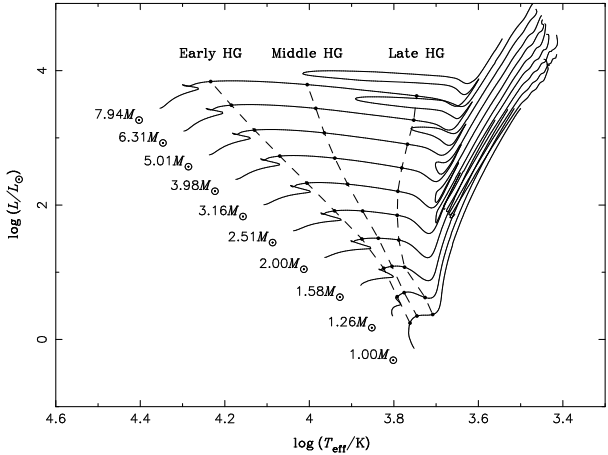}}
\caption{The evolution tracks of single stars with masses between
1 and $8M_\odot$. The three dashed lines, from left to right,
indicate the position of
an early HG, middle HG and late HG defined in this paper.
}
\label{hgrlof}
\end{figure}

We use Eggleton's \shortcite{egg71,egg72,egg73}
stellar evolution code,
which has been updated with the latest
input physics over the last 3 decades, as
described by Han et al.\ \shortcite{han94} and  Pols et al.\
\shortcite{pol95,pol98}.
The code distinguishes itself by the use of a
self-adaptive non-Lagrangian mesh, the treatment of both
convective and semiconvective mixing as a diffusion process  and the
simultaneous and implicit solution of both the stellar structure
equations and the chemical composition equations including
convective mixing. These characteristics
make the code very stable and easy to use.

The current code uses an equation of state that includes
pressure ionization and Coulomb interaction, recent
opacity tables derived from Rogers \& Iglesias \shortcite{rog92}
and Alexander \& Ferguson \shortcite{ale94a,ale94b}, nuclear reaction rates
from Caughlan \& Fowler \shortcite{cau88} and Caughlan et al.\
\shortcite{cau85}, and neutrino loss rates from Itoh et al.\
\shortcite{ito89,ito92}.

Roche lobe overflow is treated within the code. It has been
tested thoroughly and works reliably. Because the mesh-spacing
is computed along with the structure, the inclusion of
RLOF is almost trivial: just a modification of one boundary condition.
The boundary condition is written as
\begin{equation}
{{\rm d}m\over {\rm d}t}=C\cdot {\rm Max}\left[0,
({r_{\rm star}\over r_{\rm lobe}}-1)^3\right]
\label{boundary}
\end{equation}
where ${\rm d}m\over {\rm d}t$ is the mass changing rate of the star,
$r_{\rm star}$ is the radius of the star, and $r_{\rm lobe}$ the
radius of its Roche lobe. $C$ is  a constant.
With $C=1000 M_\odot/$yr, RLOF proceeds steadily, and
the lobe-filling star
overfills its Roche lobe as necessary
but never overfills its lobe by much:
$({r_{\rm star}\over r_{\rm lobe}}-1)\la 0.001$

In our calculation, we use a typical Population I (Pop I) composition
with hydrogen abundance $X=0.70$, helium abundance $Y=0.28$ and
metallicity $Z=0.02$.
We set $\alpha =l/H_{\rm p}$, the ratio of typical mixing length
to the local pressure scaleheight, to 2. Such an $\alpha$
gives a roughly correct lower main sequence, as determined observationally
by Andersen \shortcite{and91}.  It also reproduces well the location of the
red giant branch in the HR diagram for stars in the Hyades supercluster
\cite{egg85}, as determined by Bessell et al.\ \shortcite{bes89}.
A fit to the Sun also leads to $\alpha =2$ as the most
appropriate choice \cite{pol98}.

In our calculation, we just follow the evolution of the primary
(initially more massive) component in a binary system, though the code can
evolve both components quasi-simultaneously as in some previous
studies. The evolution of the primary is not
affected as long as the binary does not
come into contact.
 We assume that the RLOF is conservative in this paper.
Non-conservative cases may be studied later.

The parameter space for the model grid is three dimensional
-- primary initial mass $M_{\rm1i}$,
initial mass ratio $q_{\rm i}={M_1\over M_2}$, and primary's radius
$R_1$ at the onset of RLOF. The initial orbital period
is a function of $M_{\rm 1i}$, $q_{\rm i}$ and $R_1$.
Primary initial masses range from
1.0 to $8.0M_\odot$ at roughly equal intervals in
$\log M_{\rm 1i}$ ($M_{\rm 1i}=$ 1.0, 1.26, 1.6, 2.0, 2.5, 3.0, 4.0, 5.0,
6.3 and $8.0M_\odot$), the initial mass ratio from
1.1 to 4.0 ($q_{\rm i}=$1.1, 1.5, 2.0, 3.0, 4.0). $\log R_1$ has
3 values, $\log R_{\rm MS}+ 0.1(\log R_{\rm HG}-\log R_{\rm MS})$,
$0.5(\log R_{\rm HG}+\log R_{\rm MS})$
$\log R_{\rm HG} - 0.1(\log R_{\rm HG}-\log R_{\rm MS})$, where
$R_{\rm MS}$ and $R_{\rm HG}$ are the maximum radius
on the main sequence and in the Hertzsprung gap respectively,
for a given initial mass.
The three values correspond to onset of RLOF at early HG, middle HG
or late HG (see figure \ref{hgrlof}).
We have carried out 150 runs altogether.

\section{Results}

\begin{figure}
\centerline{\psfig{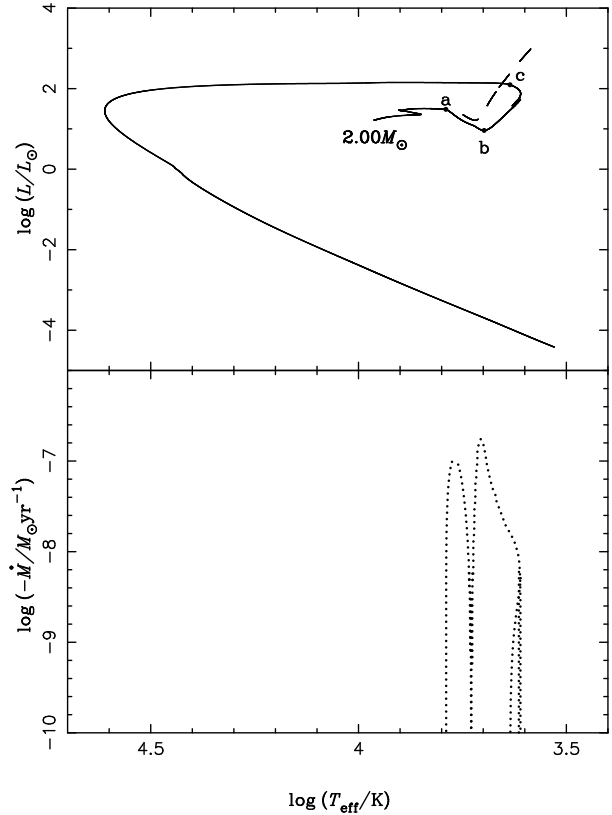}}
\caption{Evolutionary track (solid line in top panel)
of the primary of a Pop I binary system
with initial parameters 
$M_{\rm 1i}=2M_\odot$, $q_{\rm i}=M_1/ M_2=1.5$ and
$P_{\rm i}=2.551$ days (late HG). The points
a, b, and c indicate the beginning of RLOF,
the minimum luminosity during RLOF, and the end of the last episode of RLOF,
respectively.
The evolutionary track of a $2M_\odot$ Pop I single star is shown in a
dashed line for the purpose of comparison. The mass-transfer rate is
plotted in dotted line in the bottom panel.}
\label{2q15p3hrd}
\end{figure}

\begin{figure}
\centerline{\psfig{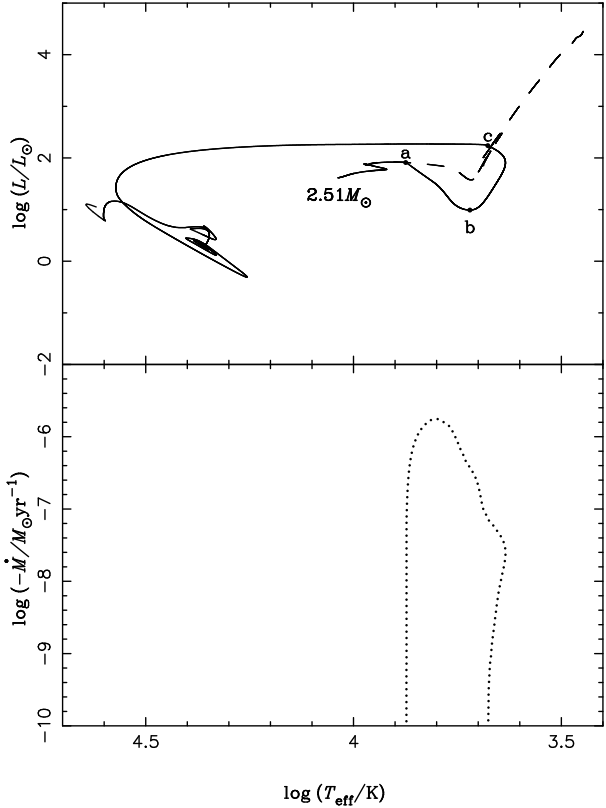}}
\caption{Similar to figure \ref{2q15p3hrd}, but
for a binary system
with initial parameters of
$M_{\rm 1i}=2.51M_\odot$, $q_{\rm i}=2$ and
$P_{\rm i}=2.559$ days (middle HG).}
\label{25q2p2hrd}
\end{figure}

\begin{figure}
\centerline{\psfig{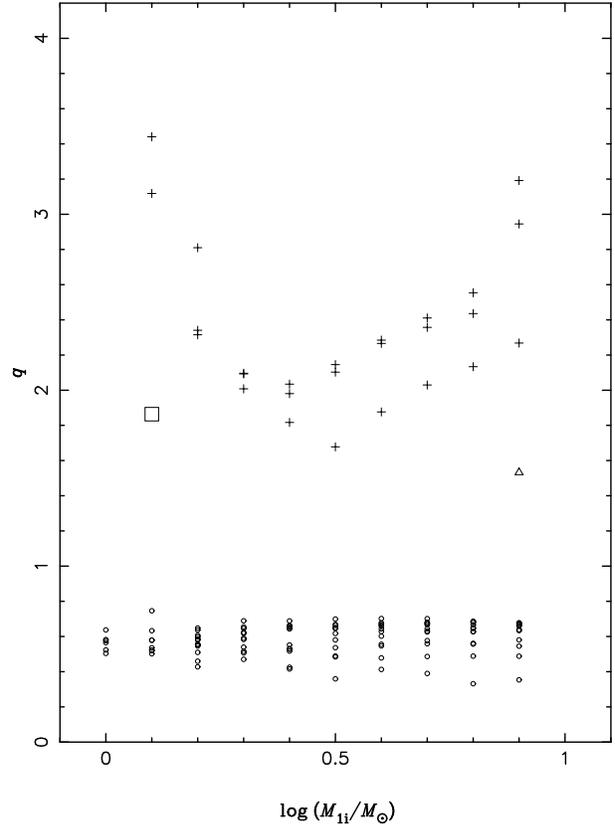}}
\caption{
Mass ratios when primaries reach the base of the red
giant branch (point b in figure \ref{2q15p3hrd} or \ref{25q2p2hrd}).
Crosses are for binaries with initial mass ratio $q_{\rm i}=4$,
circles are for binaries with initial mass ratio $q_{\rm i}=$
1.1, 1.5, 2, or 3.
The square is for
$M_{\rm 1i}=1.26M_\odot$, $q_{\rm i}=3$ and
$P_{\rm i}=0.8234$ days
and the triangle
$M_{\rm 1i}=7.94M_\odot$, $q_{\rm i}=3$ and
$P_{\rm i}=3.152$ days.
}
\label{stable}
\end{figure}

\begin{figure}
\centerline{\psfig{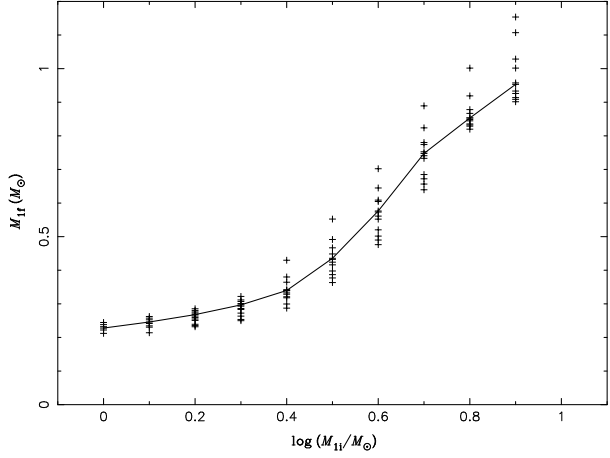}}
\caption{Remnant (WD) masses after RLOF.
The solid line is
for binaries with
initial mass ratio $q_{\rm i}=1.5$ and onset of RLOF at middle HG.
}
\label{figmif}
\end{figure}

\begin{figure}
\centerline{\psfig{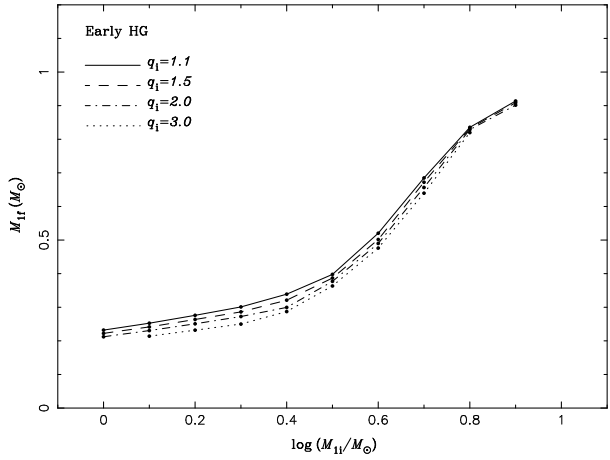}}
\caption{
Remnant (WD) masses after RLOF which begin at
early HG.
}
\label{figmifp1}
\end{figure}

\begin{figure}
\centerline{\psfig{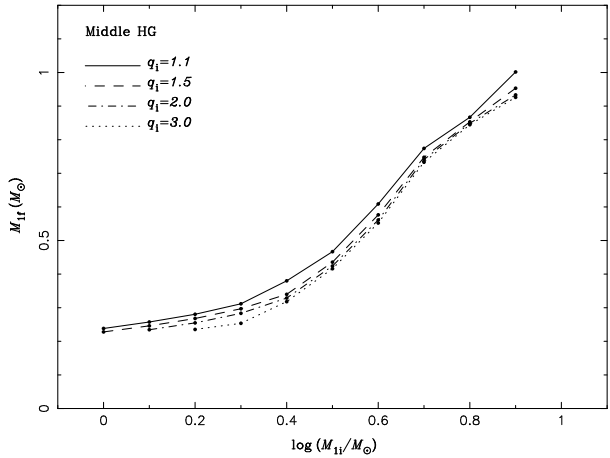}}
\caption{
Similar to figure \ref{figmifp1}, but with onset of RLOF at middle HG.
}
\label{figmifp2}
\end{figure}

\begin{figure}
\centerline{\psfig{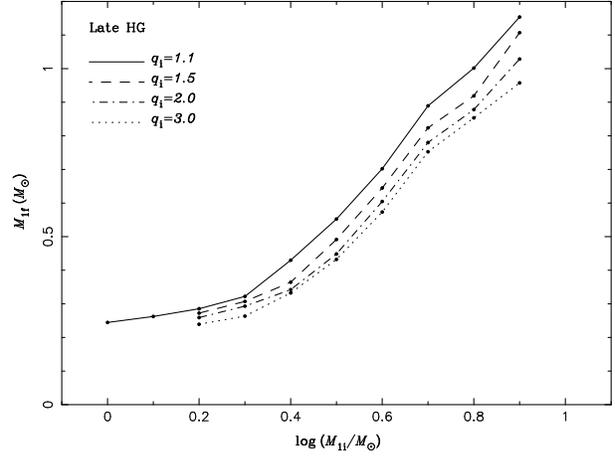}}
\caption{
Similar to figure \ref{figmifp1}, but with onset of RLOF at late HG.
}
\label{figmifp3}
\end{figure}

\begin{figure}
\centerline{\psfig{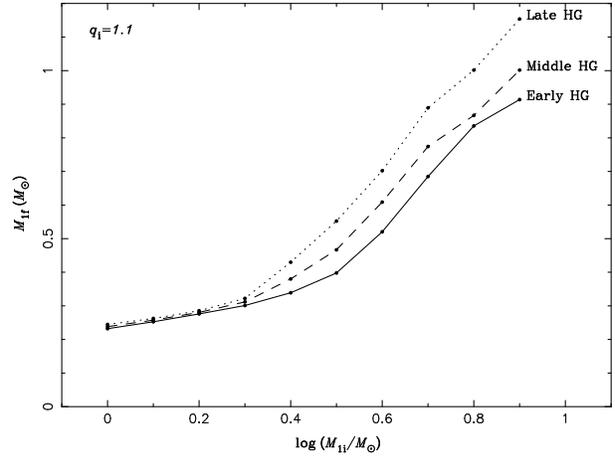}}
\caption{
Similar to figure \ref{figmifp1}, but for binaries with initial
mass ratio $q_{\rm i}=1.1$
}
\label{figmifq11}
\end{figure}

\begin{figure}
\centerline{\psfig{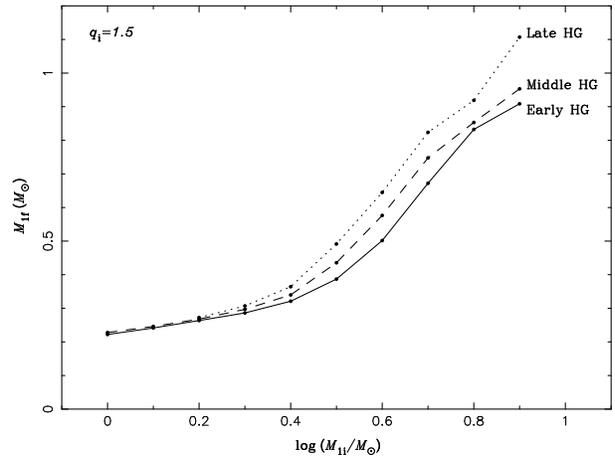}}
\caption{
Similar to figure \ref{figmifq11}, but with $q_{\rm i}=1.5$
}
\label{figmifq15}
\end{figure}

\begin{figure}
\centerline{\psfig{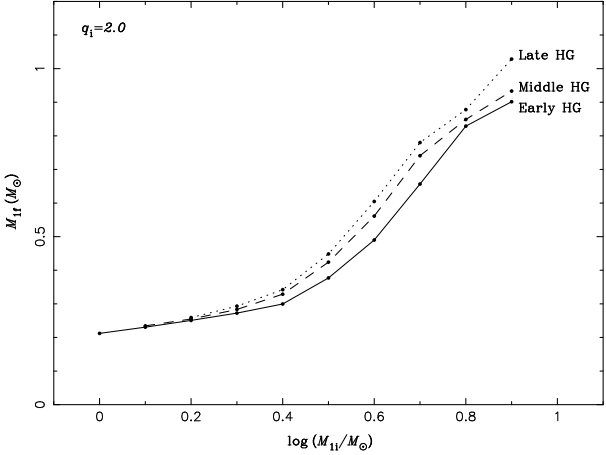}}
\caption{
Similar to figure \ref{figmifq11}, but with $q_{\rm i}=2$
}
\label{figmifq2}
\end{figure}

\begin{figure}
\centerline{\psfig{figure=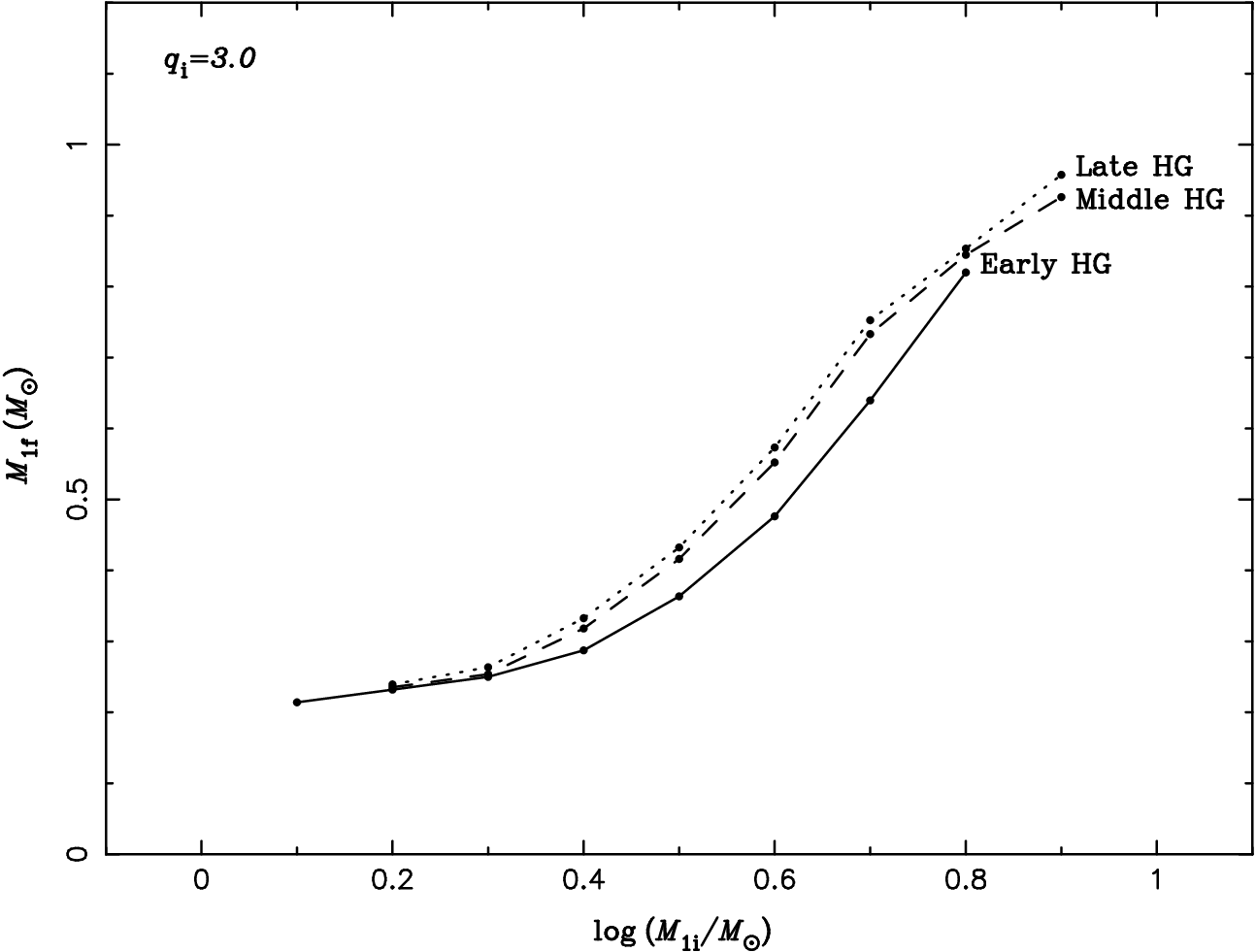,width=8cm,angle=270}}
\caption{
Similar to figure \ref{figmifq11}, but with $q_{\rm i}=3$
}
\label{figmifq3}
\end{figure}

\begin{table*}
\begin{minipage}{16cm}
\caption{Characteristics of a binary with initial parameters of
$M_{\rm 1i}=2M_\odot$, $q_{\rm i}=M_1/M_2=1.5$ and
$P_{\rm i}=2.551$ days. The events
a, b, c and d are the beginning of RLOF,
minimum luminosity during RLOF, the end of the last episode of RLOF,
and the end of the evolutionary calculation, respectively.
The time $t_1$ is the age of the primary, $M_1$ is its mass,
$\dot{M}_1$ the mass transfer rate,
$T_{\rm eff}$ its effective temperature,
$L$ its luminosity,
$R_1$ its radius,  $M_{\rm c}^{\rm He}$ its helium core mass,
$M_{\rm c}^{\rm CO}$ its carbon-oxygen core mass, and  $X_{\rm H}$ the
  hydrogen abundance at its surface.
The mass ratio of the binary is $q$ and $P$ is the orbital period.
}
\begin{tabular}{rrrrrrrrrrrr}
& $t_1$ & $M_1$ &
   $\dot{M}_1$ &
   ${\rm log}T_{\rm eff}$ &
   ${\rm log} ({L\over L_\odot})$ &
   $R_1$ & $M_{\rm c}^{\rm He}$ &
   $M_{\rm c}^{\rm CO}$ & $X_{\rm H}$ &
   $q$ & $P$ \\
& $({\rm yr})$ & $(M_\odot)$ &
   $(M_\odot {\rm yr}^{-1})$ & & &
   $(R_\odot)$ & $(M_\odot)$ &
   $(M_\odot)$ &&& $({\rm days})$ \\
a&$ 9.975  \times 10^{  8}$ &1.995&$ 0.000  $                &3.790& 1.484&
4.
853&0.211&0.000&0.700&1.500& 2.551      \\
b&$ 1.009  \times 10^{  9}$ &1.305&$-1.383  \times 10^{ -7}$ &3.698& 0.962&
4.
062&0.215&0.000&0.691&0.646& 2.601      \\
c&$ 1.079  \times 10^{  9}$ &0.307&$-1.183  \times 10^{-11}$ &3.635& 2.096&
19.
971&0.297&0.000&0.622&0.102& 59.88      \\
d&$ 1.891  \times 10^{ 10}$ &0.307&$ 0.000  $                &3.529&-4.420&
0.
018&0.306&0.000&0.622&0.102& 59.88      \\
\label{2q15p3tab}
\end{tabular}
\end{minipage}
\end{table*}

\begin{table*}
\begin{minipage}{12cm}
\caption{Coefficients for equation (\ref{mifb})}
\begin{tabular}{l|rrrrrrrrr}
  &$C_{i,1,1}$ &$C_{i,1,2}$ &$C_{i,1,3}$
  &$C_{i,2,1}$ &$C_{i,2,2}$ &$C_{i,2,3}$
  &$C_{i,3,1}$ &$C_{i,3,2}$ &$C_{i,3,3}$ \\
  $i=1$   &  1843.  &  622.1   & -249.2
          &  62.56  & -800.9   &  321.9
          & -23.28  &  173.2   & -74.99 \\
  $i=2$   &  128.9  &  247.3   & -69.61
          &  360.4  & -788.5   &  302.7
          & -131.6  &  274.4   & -101.8 \\
  $i=3$   &  24.29  &  186.7   & -34.29
          &  228.2  & -511.5   &  180.0
          & -69.89  &  163.0   & -58.72 \\
  $i=4$   &  2653.  &  173.0   &  59.91
          &  893.6  & -1099.   &  394.7
          & -490.2  &  647.2   & -202.6 \\
  $i=5$   &  28.03  &  506.9   & -97.98
          &  452.4  & -891.8   &  295.8
          & -134.7  &  280.0   & -95.06 \\
\label{coeff}
\end{tabular}
\end{minipage}
\end{table*}

Figure \ref{2q15p3hrd} shows the evolutionary track of the primary in
a Pop I binary system with initial parameters of
$M_{\rm 1i}=2M_\odot$, $q_{\rm i}=M_1/M_2=1.5$ and
$P_{\rm i}=2.551$ days.
Table \ref{2q15p3tab}
lists some characteristics for the binary at the beginning of RLOF,
at minimum luminosity during RLOF, at the end of the last episode of RLOF,
and at the end of the evolutionary calculation.
We have calculated the evolution of 150 binaries and tabulate them
in the Appendix. Figures for the
other binaries can be obtained by contacting
ZH.

As seen from figure \ref{2q15p3hrd}, there are typically three episodes 
of the RLOF. When the primary fills its Roche lobe in Hertzsprung gap, 
mass transfer begins, and the mass-transfer rate quickly
rises to $10^{-7}M_\odot {\rm yr}^{-1}$.
Mass transfer almost ceases as the primary leaves the
Hertzsprung gap but resumes with a rate of
$10^{-7}M_\odot {\rm yr}^{-1}$ when the primary evolves to the base of
red giant branch (RGB) (point b in figure \ref{2q15p3hrd}, at which
the primary has the lowest luminosity during the RLOF). The second
episode ends when
the primary starts to contract
as its envelope mass decreases after the tip of RGB, and the third episode
begins when the star expands a little more owing to a further burst of hydrogen
burning. After the third episode, the star
evolves to a helium white dwarf and cools.
 Note, however, that not all our 150 runs exhibit 3 episodes for RLOF. Some
may have only one episode (figure \ref{25q2p2hrd} for example),
and others two.

Interestingly, our calculation shows that a low mass carbon-oxygen (CO)
white dwarf (WD) may be formed as a result of Case B RLOF. Figure \ref{25q2p2hrd}
displays the evolutionary track of the primary for a binary system
with initial parameters 
$M_{\rm 1i}=2.51M_\odot$, $q_{\rm i}=M_1/M_2=2$ and
$P_{\rm i}=2.559$ days. The primary evolves to a helium white dwarf after
the RLOF. Helium is, however, ignited in the centre of the
WD as it cools
(see the loops in the left part of the top panel of figure \ref{25q2p2hrd}).
We are unable to follow the helium burning to
its end, because the code breaks down owing to the high degree of degeneracy
of the WD.
But a substantial carbon-oxygen core has developed already before the
the calculation stops. At this point we have
a white dwarf of $0.33M_\odot$ with
a carbon-oxygen core of $0.11M_\odot$.
Initial parameters of
$M_{\rm 1i}=2.51M_\odot$, $q_{\rm i}=1.1$ and
$P_{\rm i}=2.748$ days
lead to a $0.38M_\odot$ white dwarf
with a $0.21M_\odot$ CO core, but again the code breaks down
during helium burning of the WD.
A $0.43M_\odot$ white dwarf
with a $0.31M_\odot$ CO core
is obtained  from the evolution of a binary
with initial parameters of
$M_{\rm 1i}=2.51M_\odot$, $q_{\rm i}=1.1$ and
$P_{\rm i}=2.748$ days.
In this case the code completed helium burning without difficulty
so this is the true final state of the WD.

Figure \ref{stable} shows mass ratios when the primaries with different
initial parameters reach the base of
the RGB during RLOF. Almost all our 150 binaries reach the point
via stable RLOF. Only
the binaries with initial primary mass of $1M_\odot$ and initial mass
ratios $q_{\rm i} \geq 2$ experience unstable RLOF.
This is because the low-mass primaries have  convective envelopes
at the beginning of RLOF which is then dynamically unstable, and
the code breaks down. Mass ratios reach $q<0.75$ at the base of the RGB
for almost all binaries with initial primary mass
$M_{\rm 1i} \ge 1.26M_\odot$ and with initial mass ratio $q\le 3$
(see, however, the square and the triangle in figure \ref{stable} for
two exceptions). RLOF
after the base of the RGB for such binaries is stable.
Mass ratios are greater than 1.86 at the base of the RGB for all the
binaries
with initial mass ratio $q_{\rm i}=4$. RLOF on the RGB
is unstable for such binaries and the code breaks down.
As seen in figure \ref{stable}, all mass ratios are either larger than
1.5 or less than 0.75. Our calculation shows that
RLOF on the RGB is stable for binaries with $q<0.75$, and 
unstable for binaries with $q>1.5$.
In order to know the critical mass ratio
below which the RLOF on the RGB is stable
we carried out several calculations for binaries with initial primary
mass $M_{\rm 1i}=1.58M_\odot$ and with the onset of RLOF at the base
of the RGB and find that $q\le 0.8$ leads to stable RLOF while
$q\ge 0.85$ leads to unstable RLOF.

Here we are more interested in the relation between the remnant mass
of a primary and the initial parameters of the binary, because such a
relation helps us to understand the mass of white dwarf that results
from stable RLOF in the HG. Figure \ref{figmif} displays the remnant masses:
each cross is for a binary, and the solid line is for binaries with
initial mass ratio $q_{\rm i}=1.5$ and onset of RLOF at middle HG.
We see that the masses scatter around the solid line by $\pm 22$ per cent.
This means that the remnant (WD) mass is dependent on the initial mass
ratio $q_{\rm i}$ and the primary's radius $R_1$ at the onset of RLOF
(or the initial orbital period $P_{\rm i}$), as well as on the initial mass
$M_{\rm 1i}$ of the primary.

As shown by figures \ref{figmifp1}, \ref{figmifp2} and \ref{figmifp3},
the remnant mass gets smaller if the initial mass ratio is bigger.
This is because a high mass ratio leads to a high mass-transfer
rate, and RLOF therefore takes a shorter time and gives a smaller
core mass
when the primary's envelope is totally stripped off.

As seen from figures \ref{figmifq11}, \ref{figmifq15},
\ref{figmifq2} and \ref{figmifq3}, the remnant mass gets larger
if RLOF begins later in the HG. This is simply because
the core mass grows as the primary evolves in the HG.

For $q_{\rm i}=1.5$ and middle-HG RLOF, the remnant mass $M_{\rm 1f}$
can be fitted, with an error of 1.3 per cent for
$1M_\odot \le M_{\rm 1i} \le 8M_\odot$ by
\begin{equation}
m_{\rm 1f}={174.6-36m_{\rm 1i}^2+19m_{\rm 1i}^{2.5} \over
             1000-369m_{\rm 1i}+52m_{\rm 1i}^2},
\label{mifqp}
\end{equation}
where $m_{\rm 1f}=M_{\rm 1f}/M_\odot$ and $m_{\rm 1i}=M_{\rm 1i}/M_\odot$.
We have found a more general fit for the remnant mass
as a function of $M_{\rm 1i}$,
$q_{\rm i}$ and $R_1$ but with an error of 3.6 per cent
for $1M_\odot \le M_{\rm 1i} \le 8M_\odot$ and
$1.1 \le q_{\rm i} \le 3$:
\begin{equation}
m_{\rm 1f}={A_1-A_2m_{\rm 1i}^2+A_3m_{\rm 1i}^{2.5} \over
            10000-A_4m_{\rm 1i}+A_5m_{\rm 1i}^2},
\label{mif}
\end{equation}
where
\begin{equation}
         A_i=B_{i,1}+B_{i,2}q_{\rm i}+B_{i,3}q_{\rm i}^2
\label{mifa}
\end{equation}
and
\begin{equation}
  B_{i,j}=C_{i,j,1}+C_{i,j,2}D+C_{i,j,3}D^2.
\label{mifb}
\end{equation}
The coefficients $C_{i,j,k} (i=1,5;j=1,3;k=1,3)$ are given in
table \ref{coeff}, and $D$ is defined below.
$D\approx$1, 2, and 3 correspond to early-HG, middle-HG and
 late-HG RLOF (see figure \ref{hgrlof}):
\begin{equation}
      D={2\log (R_1/R_\odot) - 3D_1+D_2 \over D_2-D_1}
\label{mifr}
\end{equation}
where
\begin{equation}
      D_1={0.121+4.16 \log m_{\rm 1i} \over
         1+3.16 (\log m_{\rm 1i})^{0.5}}
\label{mifr1}
\end{equation}
and
\begin{equation}
      D_2=0.285 + 1.06 \log m_{\rm 1i} + 0.822 (\log m_{\rm 1i})^2.
\label{mifr2}
\end{equation}

\section{Discussion}

Van der Linden \shortcite{van87}
and de Loore \& Vanbeveren \shortcite{del95} have calculated
conservative Case-B evolution of intermediate mass binaries,
though only for one or two mass ratios. However, it is not possible
to compare our results with theirs in a one-to-one way, as none of
their binaries has exactly
the same initial parameters (primary mass, mass ratio
and orbital period) as any of ours. But we do make a comparison
and find that our results agree with theirs reasonably for
binaries with initial primary mass less than $4M_\odot$. The difference
between our results and theirs is small and is mainly
due to the following differences between our work and theirs.
We use the OPAL opacity tables \cite{rog92} while
van der Linden used the LAOS ones. The treatment of RLOF in
our code is different from theirs. Ours is more efficient in
computation, but theirs may be more realistic physically. We
do not include convective core overshooting but de Loore
\& Vanbeveren \shortcite{del95} do. And, perhaps more importantly,
our stellar evolution code is different from theirs.

For binaries with initial primary mass larger than $5M_\odot$,
there exist very large differences between our results and
theirs. For example, the hydrogen abundance at the primary's surface
is almost zero after RLOF in our model for
a binary with initial parameters
of $M_{\rm 1i}=5M_\odot$, $q_{\rm i}=1.1$ and
$P_{\rm i}=22.52$ days, while the hydrogen abundance
of a similar binary in their model is about 0.2.
The timescale of RLOF in our model is much larger than that in
their model. This is explained as follows. For a binary with its
primary initially more massive than $5M_\odot$, the primary
fills its Roche lobe in the Hertzsprung gap and RLOF begins.
The primary becomes a red giant during the RLOF, and its envelope
is stripped away gradually. When the envelope is very small, the
primary contracts and becomes a helium star (but with a hydrogen-helium
envelope). Helium is ignited in the centre and the star expands
after central helium is exhausted. The star becomes a red giant again
and fills its Roche lobe and the second episode of RLOF begins.
At the end of the second episode of RLOF, the surface hydrogen abundance
becomes zero obviously.
Van der Linden \shortcite{van87}
and de Loore \& Vanbeveren \shortcite{del95} have only followed
the first episode of RLOF and they stop the evolution
when the central helium is exhausted. We have followed the two
episodes and our calculation is terminated when the primary
becomes a helium WD or carbon-oxygen WD in most cases.
Therefore the timescale of RLOF in
our model (the time between the onset of the first episode of RLOF
and the end of the last episode by our definition) is much larger than
that of van der Linden \shortcite{van87} and de Loore \& Vanbeveren
\shortcite{van87}.

Most of the double degenerates observed by Marsh's group
\cite{mar00} have primary masses less than $0.5M_\odot$ and they
are thought to be helium WDs. Our calculation shows that
a low mass (as low as $0.33M_\odot$) carbon-oxygen WD may be formed from
stable
RLOF in HG, so that some of the presumed helium WDs may actually be
carbon-oxygen WDs.

In order to get better consistency with observations,
Iben et al.\  \shortcite{ibe97} use a rather large common-envelope
ejection efficiency in their binary population synthesis studies.
They assume that RLOF in the HG is always dynamically unstable and that
RLOF leads to the formation of a common envelope. Our calculation shows
however that RLOF is very likely to be stable, even if the initial
mass ratio is as large as 3. If the initial mass ratio is even larger,
RLOF is stable at the beginning and part of the primary's
envelope is transferred to the secondary before the RLOF becomes
dynamically unstable. If the process of RLOF is treated as common-envelope
evolution, as by Iben et al., a high efficiency is required for
common-envelope ejection.

Most of the previous studies treated binaries with only one or two
initial mass ratios and with the onset of RLOF early in the HG only.
Our study is more comprehensive in that our calculation is
for a rather complete space of initial parameters. It has been believed that
the WD mass derived from the RLOF does not depend much on the initial
mass ratio or the initial orbital period \cite{deg93}. However,
our investigation shows that the dependence is quite significant. Therefore
the binary evolutionary models in this paper should be included in
any realistic binary population synthesis model.

To obtain a copy of all the evolutionary tracks (similar to
figure \ref{2q15p3hrd}), all tables for characteristics of RLOF
(similar to table \ref{2q15p3tab}), and equations
\ref{mif} to \ref{mifr2} in FORTRAN code, please send an email
to ZH who will provide them via ftp.

\section*{acknowledgments}
ZH thanks the Institute of Astronomy, Cambridge for its hospitality
and financial support, the Royal Society for a fellowship and
the Isaac Newton Trust for kind support. ZH also thanks the support
from the Chinese Natural Science Foundation (Grant No. 19925312) and from
the
973 scheme. CAT is very grateful to PPARC for an advanced fellowship.
We would like to thank Philipp Podsiadlowski for valuable
discussions and suggestions, and thank the referee, Prof. D. Vanbeveren,
for useful suggestions.

\appendix
\section{RLOF for all the binaries}

Table A1 lists all the binaries in the manner of table 1
in the body of the paper:
\begin{description}
\item a: the onset of RLOF
\item b: the minimum luminosity during RLOF (i.e. just reaching the RGB)
\item c: at the end of RLOF
         (RLOF may have several episodes, there is no RLOF any more after c)
\item d: at the end of calculation (some of the calculations break
         during central helium ignition)
\end{description}
We usually list stellar parameters at a, b, c and d. However,
the code breaks down
when RLOF is unstable. In that case, we only list parameters at a
if RLOF is unstable at onset, or at a and b
if RLOF is stable at the onset but becomes unstable after the primary 
reaches the RGB.

\end{document}


\begin{appendix}

\centerline{{\bf Table A1:} RLOF in Hertzsprung gap}
\scriptsize
\begin{longtable}{rrrrrrrrrrrrr}
 & $\scriptstyle t_1$ & $\scriptstyle M_1$ &
   $\scriptstyle \dot{M}_1$ &
   $\scriptstyle {\rm log}T_{\rm eff}$ & 
   $\scriptstyle {\rm log} ({L\over L_\odot})$ &
   $\scriptstyle R_1$ & $\scriptstyle M_{\rm c}^{\rm He}$ & 
   $\scriptstyle M_{\rm c}^{\rm CO}$ & $\scriptstyle X_{\rm H}$ & 
   $\scriptstyle q$ & $\scriptstyle P$ \\
 & $\scriptstyle ({\rm yr})$ & $\scriptstyle (M_\odot)$ &
   $\scriptstyle (M_\odot {\rm yr}^{-1})$ & & & 
   $\scriptstyle (R_\odot)$ & $\scriptstyle (M_\odot)$ & 
   $\scriptstyle (M_\odot)$ &&& $\scriptstyle ({\rm days})$ \\ 
\endfirsthead
\caption[]{continued}\\
 & $\scriptstyle t_1$ & $\scriptstyle M_1$ &
   $\scriptstyle \dot{M}_1$ &
   $\scriptstyle {\rm log}T_{\rm eff}$ & 
   $\scriptstyle {\rm log} ({L\over L_\odot})$ &
   $\scriptstyle R_1$ & $\scriptstyle M_{\rm c}^{\rm He}$ & 
   $\scriptstyle M_{\rm c}^{\rm CO}$ & $\scriptstyle X_{\rm H}$ & 
   $\scriptstyle q$ & $\scriptstyle P$ \\
 & $\scriptstyle ({\rm yr})$ & $\scriptstyle (M_\odot)$ &
   $\scriptstyle (M_\odot {\rm yr}^{-1})$ & & & 
   $\scriptstyle (R_\odot)$ & $\scriptstyle (M_\odot)$ & 
   $\scriptstyle (M_\odot)$ &&& $\scriptstyle ({\rm days})$ \\ 
\endhead
\multicolumn{13}{r}{continued on the next page}
\endfoot
\endlastfoot
&&&&&&&&&&&\\
a&$ 9.549  \times 10^{  9}$ &1.000&$ 0.000  $                &3.757& 0.222&   1.322&0.040&0.000&0.700&1.100& 0.5296     \\
b&$ 1.202  \times 10^{ 10}$ &0.640&$-2.328  \times 10^{-10}$ &3.677&-0.061&   1.376&0.107&0.000&0.687&0.504& 0.7427     \\
c&$ 1.379  \times 10^{ 10}$ &0.232&$-2.589  \times 10^{-13}$ &3.710& 1.045&   4.231&0.218&0.000&0.616&0.138& 6.747      \\
d&$ 3.608  \times 10^{ 10}$ &0.232&$ 0.000  $                &3.500&-4.424&   0.020&0.230&0.000&0.616&0.138& 6.747      \\
&&&&&&&&&&&\\
a&$ 1.081  \times 10^{ 10}$ &1.000&$ 0.000  $                &3.742& 0.332&   1.607&0.094&0.000&0.700&1.100& 0.7121     \\
b&$ 1.136  \times 10^{ 10}$ &0.690&$-5.770  \times 10^{-10}$ &3.677& 0.075&   1.609&0.119&0.000&0.687&0.565& 0.8995     \\
\end{longtable}

\end{appendix}